\documentclass[twocolumn]{article}

\usepackage{graphics,graphicx,tikz,algpseudocode,algorithm, siunitx, amsfonts}
\usepackage{authblk}
\usepackage{xcolor,colortbl}
\usepackage{mathtools}
\usepackage{authblk} % For author affiliation
\usepackage{amsmath}
\usetikzlibrary{shapes.geometric}
\usepackage[margin=1in]{geometry}
\PassOptionsToPackage{hyphens}{url}\usepackage{hyperref}

\usepackage{subcaption}
\newtheorem{definition}{Definition}
\usepackage{xargs}
\usepackage{xspace}
\usepackage[colorinlistoftodos,prependcaption,textsize=small]{todonotes}

\DeclarePairedDelimiter{\abs}{\lvert}{\rvert}

\def\LittleKtwo{\tikz[baseline={([yshift=-0.7ex]current bounding box.center)}]{
		\fill (0,0) circle (1.5pt) coordinate (A);
		\fill (0,4ex) circle (1.5pt) coordinate (B);
		\draw (A)--(B);}
}

\def\LittleKtwoComp{\tikz[baseline={([yshift=-0.7ex]current bounding box.center)}]{
		\fill (0,0) circle (1.5pt) coordinate (A);
		\fill (0,4ex) circle (1.5pt) coordinate (B);}
}

\def\Ktwo{\tikz[baseline={([yshift=-0.7ex]current bounding box.center)}]{
		\fill (0,0) circle (1.5pt) coordinate (A);
		\fill (0,6ex) circle (1.5pt) coordinate (B);
		\draw (A)--(B);}
}

\def\KthreeComp{\tikz[baseline={([yshift=-1ex]current bounding box.center)}]{
		\fill (0,0) circle (1.5pt) coordinate (A);
		\fill (6ex,0) circle (1.5pt) coordinate (C);
		\fill (3ex,5.196ex) circle (1.5pt) coordinate (B);}
}

\def\PthreeComp{\tikz[baseline={([yshift=-1ex]current bounding box.center)}]{
		\fill (0,0) circle (1.5pt) coordinate (A);
		\fill (6ex,0) circle (1.5pt) coordinate (C);
		\fill (3ex,5.196ex) circle (1.5pt) coordinate (B);
         \draw (A)--(B);}
}

\def\Pthree{\tikz[baseline={([yshift=-1ex]current bounding box.center)}]{
		\fill (0,0) circle (1.5pt) coordinate (A);
		\fill (6ex,0) circle (1.5pt) coordinate (C);
		\fill (3ex,5.196ex) circle (1.5pt) coordinate (B);
         \draw (B)--(C)--(A);}
}

\def\Kthree{\tikz[baseline={([yshift=-1ex]current bounding box.center)}]{
		\fill (0,0) circle (1.5pt) coordinate (A);
		\fill (6ex,0) circle (1.5pt) coordinate (C);
		\fill (3ex,5.196ex) circle (1.5pt) coordinate (B);
              \draw (A)--(B)--(C)--(A);}
}

\def\KfourComp{\tikz[baseline={([yshift=-0.7ex]current bounding box.center)}]{
		\fill (0,0) circle (1.5pt) coordinate (A);
		\fill (0,6ex) circle (1.5pt) coordinate (B);
      \fill (6ex,6ex) circle (1.5pt) coordinate (C);
      \fill (6ex,0) circle (1.5pt) coordinate (D);
		}
}

\def\evv{\tikz[baseline={([yshift=-0.7ex]current bounding box.center)}]{
		\fill (0,0) circle (1.5pt) coordinate (A);
		\fill (0,6ex) circle (1.5pt) coordinate (B);
      \fill (6ex,6ex) circle (1.5pt) coordinate (C);
      \fill (6ex,0) circle (1.5pt) coordinate (D);
		\draw (A)--(B);}
}

\def\ee{\tikz[baseline={([yshift=-0.7ex]current bounding box.center)}]{
		\fill (0,0) circle (1.5pt) coordinate (A);
		\fill (0,6ex) circle (1.5pt) coordinate (B);
      \fill (6ex,6ex) circle (1.5pt) coordinate (C);
      \fill (6ex,0) circle (1.5pt) coordinate (D);
		\draw (A)--(B) (C)--(D);}
}

\def\Pthreev{\tikz[baseline={([yshift=-0.7ex]current bounding box.center)}]{
		\fill (0,0) circle (1.5pt) coordinate (A);
		\fill (0,6ex) circle (1.5pt) coordinate (B);
      \fill (6ex,6ex) circle (1.5pt) coordinate (C);
      \fill (6ex,0) circle (1.5pt) coordinate (D);
		\draw (B)--(A)--(D);}
}

\def\Kthreev{\tikz[baseline={([yshift=-0.7ex]current bounding box.center)}]{
		\fill (0,0) circle (1.5pt) coordinate (A);
		\fill (0,6ex) circle (1.5pt) coordinate (B);
      \fill (6ex,6ex) circle (1.5pt) coordinate (C);
      \fill (6ex,0) circle (1.5pt) coordinate (D);
		\draw (B)--(A)--(D)--(B);}
}

\def\Konethree{\tikz[baseline={([yshift=-0.7ex]current bounding box.center)}]{
		\fill (0,0) circle (1.5pt) coordinate (A);
		\fill (0,6ex) circle (1.5pt) coordinate (B);
      \fill (6ex,6ex) circle (1.5pt) coordinate (C);
      \fill (6ex,0) circle (1.5pt) coordinate (D);
		\draw (B)--(A)--(D) (A)--(C);}
}

\def\Pfour{\tikz[baseline={([yshift=-0.7ex]current bounding box.center)}]{
		\fill (0,0) circle (1.5pt) coordinate (A);
		\fill (0,6ex) circle (1.5pt) coordinate (B);
      \fill (6ex,6ex) circle (1.5pt) coordinate (C);
      \fill (6ex,0) circle (1.5pt) coordinate (D);
		\draw (B)--(A)--(D)--(C);}
}

\def\Kthreep{\tikz[baseline={([yshift=-0.7ex]current bounding box.center)}]{
		\fill (0,0) circle (1.5pt) coordinate (A);
		\fill (0,6ex) circle (1.5pt) coordinate (B);
      \fill (6ex,6ex) circle (1.5pt) coordinate (C);
      \fill (6ex,0) circle (1.5pt) coordinate (D);
		\draw (B)--(A)--(D)--(B) (D)--(C);}
}

\def\Cfour{\tikz[baseline={([yshift=-0.7ex]current bounding box.center)}]{
		\fill (0,0) circle (1.5pt) coordinate (A);
		\fill (0,6ex) circle (1.5pt) coordinate (B);
      \fill (6ex,6ex) circle (1.5pt) coordinate (C);
      \fill (6ex,0) circle (1.5pt) coordinate (D);
		\draw (B)--(A)--(D)--(C)--(B);}
}

\def\Kfourme{\tikz[baseline={([yshift=-0.7ex]current bounding box.center)}]{
		\fill (0,0) circle (1.5pt) coordinate (A);
		\fill (0,6ex) circle (1.5pt) coordinate (B);
      \fill (6ex,6ex) circle (1.5pt) coordinate (C);
      \fill (6ex,0) circle (1.5pt) coordinate (D);
		\draw (B)--(A)--(D)--(C)--(B)--(D);}
}

\def\Kfour{\tikz[baseline={([yshift=-0.7ex]current bounding box.center)}]{
		\fill (0,0) circle (1.5pt) coordinate (A);
		\fill (0,6ex) circle (1.5pt) coordinate (B);
      \fill (6ex,6ex) circle (1.5pt) coordinate (C);
      \fill (6ex,0) circle (1.5pt) coordinate (D);
		\draw (B)--(A)--(D)--(C)--(B)--(D) (A)--(C);}
}

\begin{document}
\title{ Tight Practical Bounds for Subgraph Densities in Ego-centric Networks}

%%
%% The "author" command and its associated commands are used to define the authors and their affiliations.
\author[1]{Connor Mattes\thanks{Email: clmatte@sandia.gov }}
\author[1]{Esha Datta\thanks{Email: edatta@sandia.gov \newline Sandia National Laboratories is a multimission laboratory managed and operated by National Technology and Engineering Solutions of Sandia, LLC, a wholly owned subsidiary of Honeywell International, Inc., for the U.S. Department of Energy’s National Nuclear Security Administration under contract DE-NA-0003525.}}
\author[2]{Ali Pinar\thanks{Email: alipinar@gmail.com \newline This work was completed while Ali Pinar was at Sandia National Laboratories.}}

\affil[1]{Sandia National Laboratories, 7011 East Avenue, Livermore, CA 94550}
\maketitle

\begin{abstract}
   Subgraph densities play a crucial role in network analysis, especially for the identification and interpretation of meaningful substructures in complex graphs. 
   For instance, triangle densities provide a measure of cohesiveness of a graph and are used to distinguish between social networks and other types of graphs. 
   Localized subgraph densities, in particular, can provide valuable insights into graph structures.

   Distinguishing between mathematically-\allowbreak determined and domain-\allowbreak driven subgraph density features, however, poses challenges. 
   For instance, the lack or presence of certain structures can be directly explained by graph density or degree distribution. 
   These differences are especially meaningful in applied contexts as they allow us to identify instances where the data induces specific network structures, such as friendships in social networks.

   The goal of this paper  is to measure these differences across various types of graphs, conducting social media analysis from a network perspective.
   To this end, we first provide tighter bounds on subgraph densities. We then introduce the \emph{subgraph spread ratio} to quantify the realized subgraph densities of specific networks relative to the feasible bounds. 
   Our novel approach combines techniques from flag algebras, motif-counting, and topological data analysis. 
   Crucially, effective adoption of the state-of-the-art in the plain flag algebra method yields feasible regions up to three times tighter than prior best-known results, thereby enabling more accurate and direct comparisons across graphs.

   We additionally perform an empirical analysis of 11 real-world networks. 
   We observe that social networks consistently have smaller subgraph spread ratios than other types of networks, such as linkage-mapping networks for Wikipedia pages. 
   This aligns with our intuition about social relationships: such networks have meaningful structure that makes them distinct. 
   The subgraph spread ratio enables the quantification of intuitive understandings of network structures and provides a metric for comparing types of networks. 
   We provide C++ code for computing the subgraph spread ratio at \url{https://github.com/sandialabs/Tight-Practical-Bounds-for-Subgraph-Densities-in-Ego-centric-Networks}.
\end{abstract}

\section{Introduction}
\begin{figure}
       \includegraphics[scale=0.5]{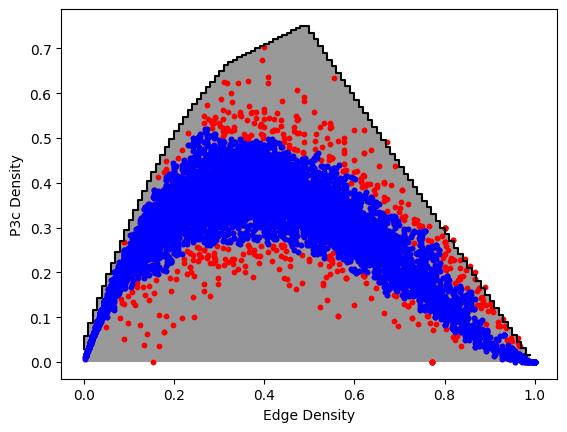}
       \caption{Localized point cloud with respect to the graph on 3 vertices with a single edge ($\overline{P_3}$) of some subset of Facebook pages \cite{MUSAE}.
       The red points are outliers, consisting of $5\%$ of the data, determined by our pruning algorithm. 
       The gray shaded area is the feasible region of values given by the plain flag algebra method.
       The subgraph spread ratio is $0.393$, meaning that roughly $40\%$ of the feasible region is covered by the localized point cloud.}
       \label{algorithmExample}
\end{figure}

Subgraph densities are a crucial tool in the field of network analysis, particularly for identifying and interpreting meaningful substructures in complex graphs. 
For instance, subgraph densities have proven essential in applications such as network classification \cite{classify} and community detection \cite{community}.
Induced subgraphs also play a pivotal role in theoretic settings, such as the fields of extremal graph theory \cite{lovasz} and graph coloring \cite{coloring}. 

Many applications, such as frequent subgraph mining \cite{fsm}, study subgraph densities over the entire graph. 
Our analysis instead focuses on local subgraph densities in a vertex's neighborhood.
These are referred to in the literature as ego-centric networks \cite{ego1,ego2,ego3}. 
The focus on ego-centric networks represents a paradigm shift from the typical approach: rather than studying one large, possibly sparse graph, we instead focus on many smaller and denser graphs. 
An exemplar of this perspective is \cite{main}, in which the authors study structural graph properties of a subset of Facebook pages and users by plotting the edge density in each vertex's neighborhood against the density of some other graph, $H$. 
\begin{definition}
We define the \textit{localized point cloud with respect to} $\mathit{H}$ as the scatter plot in $\mathbb{R}^2$ plotting $K_2$ density on the x-axis against $H$ density on the y-axis.
When the graph $H$ is apparent we will refer simply to the \emph{localized point cloud}. 
\end{definition}
Figure \ref{algorithmExample} shows a visualization of a localized point cloud as a scatter plot.
This visualization aids in understanding the graph's structure at a local level and will allow us to apply techniques from topological data analysis to the problem.

The localized point cloud can provide unexpected insights into subgraph structure and, therefore, the dataset that generates the graph model. 
For instance, we might expect the localized point cloud of a sufficiently large graph to realize most feasible values. 
This intuition holds for many arbitrary graphs, but not for social networks. 
We observe that social networks induce localized point clouds that occupy a strict subset of the feasible region (excluding a few outlying data points). 
This unexpected phenomenon holds across several datasets and types of social networks; that is, the graphs produced by social network data are qualitatively distinct from arbitrary graphs of the same size and this distinction can be observed through the localized point cloud. 
We call graphs that occupy a smaller portion of the feasible region more \emph{domain-driven} as they are strongly restricted, but not because of the feasible region.
This is opposed to \emph{mathematically-determined} graphs whose structures are constrained by the feasible region, as they occupy a large portion of it. 
Figure \ref{algorithmExample} contains domain-driven structure demonstrated by the distinct shape of the localized point cloud.

We note that mathematically-determined and domain-driven correspond, loosely, to the ideas of exogenity and endogenity respectively \cite{exen1, exen2, ERGM}. 
Our work is not a critique of these past methods, but instead a novel approach to the analysis of subgraph features specific to ego-centric networks, and a new way of thinking about these ideas.

This work seeks to characterize and understand the distinct patterns in the localized point clouds of certain graphs. 
Our approach is two-fold: 
\begin{itemize}
   \item To determine accurate feasible regions for localized point clouds, and 
   \item To quantify the spread in local subgraph structure.
\end{itemize} 

\begin{figure}
   \includegraphics[scale=0.5]{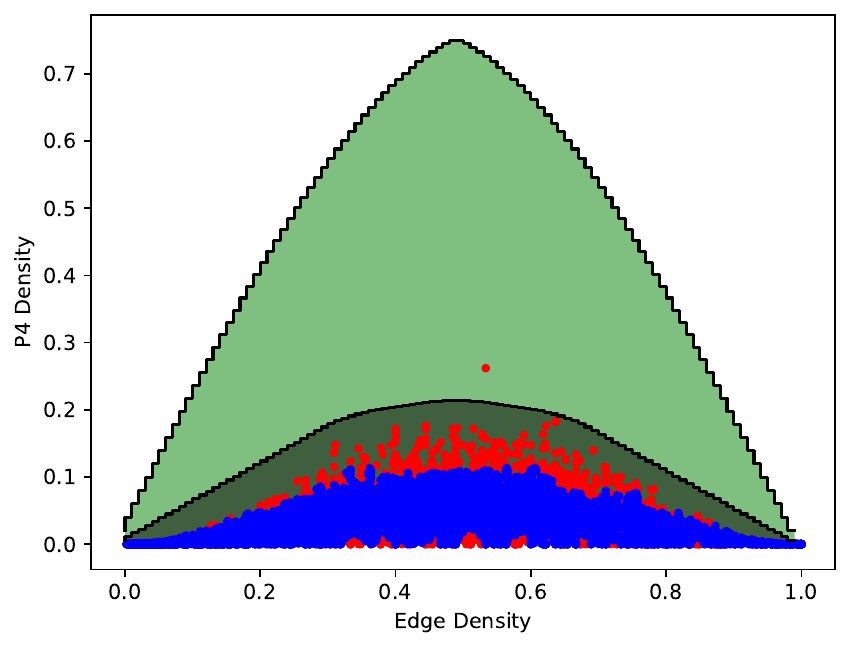}
   \caption{Localized point cloud with respect to the path graph on four vertices ($P_4$) of some subset of Facebook pages \cite{MUSAE}.
   The red points are outliers, determined by our pruning algorithm. 
   The darker shaded area is the feasible region of values given by the plain flag algebra method, whereas the lighter green region is the feasible region given in \cite{main}.
   Our new feasible region is less than a third of the size.
   Note that these feasible regions only hold asymptotically in the size of the ego-centric networks, which accounts for the data-point that falls outside of our bound.}
   \label{comparison}
\end{figure}

We leverage the plain flag algebra method \cite{flag} to tighten the previous best bounds \cite{main} on the feasible region of localized point clouds. 
The feasible region is, in some cases, more than three times smaller than previously thought, see Figure \ref{comparison}. 
We provide a more detailed analysis of these feasible regions in Section \ref{sec:results}.

We then introduce the subgraph spread ratio (of $H$), which we define as the ratio of the area of the localized point cloud (with respect to $ H$) to the area of the feasible region. 
Subgraph spread ratios near 1 are indicative of mathematically-determined structure, while values close to 0 indicate domain-driven structure. 
For example, Figure \ref{algorithmExample} shows a localized point cloud with a subgraph spread ratio of $\approx 0.4$, indicating that only $40\%$ of the feasible region is covered by data points. 
We calculate subgraph spread ratios for various large real-world networks and demonstrate that the average subgraph spread ratio is consistently smaller for social networks than for other graphs.

 The contributions of this paper are:
 \begin{itemize}
 \item Novel use of flag algebras.
 \item Conception of subgraph spread ratio as meaningful metric.
 \item Description of algorithm to calculate the subgraph spread ratio.
 \item Study of the subgraph spread ratio for real-world networks.
 \item Inclusion of code for calculating the subgraph spread ratio. 
 \end{itemize}

Our paper is organized into the following sections: 
Section \ref{sec:math} provides the mathematical background necessary to discuss localized point clouds, including an introduction to the plain flag algebra method.
Section \ref{alg} formally defines the subgraph spread ratio and provides the algorithm to calculate it. 
In Section \ref{sec:results}, we analyze the accuracy of the plain flag algebra method and the predictive power of the subgraph spread ratio. 
Section \ref{sec:future} provides a list of future directions for this line of work,  and Section \ref{sec:conclusion} summarizes our results and their impacts.

\section{Preliminaries}
\label{sec:math}

This section outlines key terminology, notation, and the plain flag algebra method. 

\subsection*{Graph Theory Background}
\begin{figure}[hpb]
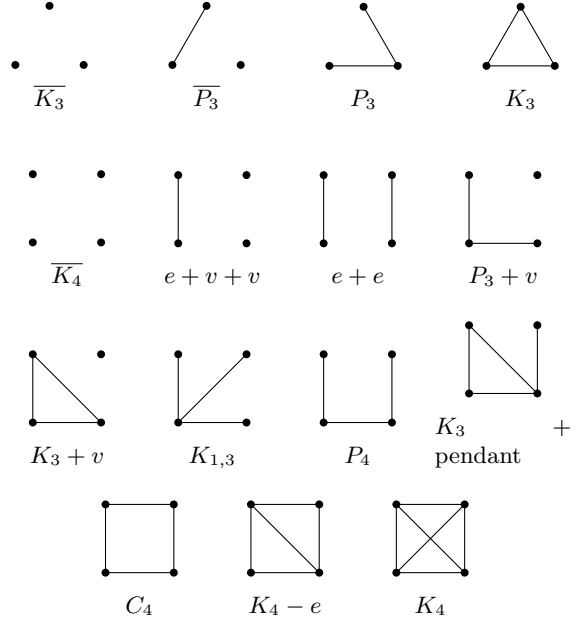

   \centering
   \begin{subfigure}[b]{0.11\textwidth}
       \centering
       \KthreeComp
       \caption*{$\overline{K_3}$}
   \end{subfigure}
   \hfill
   \begin{subfigure}[b]{0.11\textwidth}
       \centering
       \PthreeComp
       \caption*{$\overline{P_3}$}
   \end{subfigure}
   \hfill
   \begin{subfigure}[b]{0.11\textwidth}
       \centering
       \Pthree
       \caption*{$P_3$}
   \end{subfigure}
   \hfill
   \begin{subfigure}[b]{0.11\textwidth}
       \centering
       \Kthree
       \caption*{$K_3$}
   \end{subfigure}

   \vspace{20pt}

   \begin{subfigure}[b]{0.11\textwidth}
      \centering
      \KfourComp
      \caption*{$\overline{K_4}$}
   \end{subfigure}
   \begin{subfigure}[b]{0.11\textwidth}
      \centering
      \evv
      \caption*{$e+v+v$}
   \end{subfigure}
   \begin{subfigure}[b]{0.11\textwidth}
      \centering
      \ee
      \caption*{$e+e$}
   \end{subfigure}
   \begin{subfigure}[b]{0.11\textwidth}
      \centering
      \Pthreev
      \caption*{$P_3+v$}
   \end{subfigure}

   \vspace{10pt}

   \begin{subfigure}[b]{0.11\textwidth}
      \centering
      \Kthreev
      \caption*{$K_3+v$}
   \end{subfigure}
   \begin{subfigure}[b]{0.11\textwidth}
      \centering
      \Konethree
      \caption*{$K_{1,3}$}
   \end{subfigure}
   \begin{subfigure}[b]{0.11\textwidth}
      \centering
      \Pfour
      \caption*{$P_4$}
   \end{subfigure}
   \begin{subfigure}[b]{0.11\textwidth}
      \centering
      \Kthreep
      \caption*{$K_3+\text{pendant}$}
   \end{subfigure}

   \vspace{10pt}

   \begin{subfigure}[b]{0.11\textwidth}
      \centering
      \Cfour
      \caption*{$C_4$}
   \end{subfigure}
   \begin{subfigure}[b]{0.11\textwidth}
      \centering
      \Kfourme
      \caption*{$K_4-e$}
   \end{subfigure}
   \begin{subfigure}[b]{0.11\textwidth}
      \centering
      \Kfour
      \caption*{$K_4$}
   \end{subfigure}
   \caption{All graphs on three and four vertices} 
   \label{fig::graphs}
\end{figure}

Let $G = (V,E)$ be a simple graph. 
The vertex set and edge set of the graph are referred to as $V(G)$ and $E(G)$, respectively. 
For any vertex $v$ in $V(G)$, let $N(v) = \{u \in V(G) : uv \in E(G) \}$ denote its neighborhood. 
Note that $v$ itself is not included in $N(v)$. 
The degree of vertex $v$ is defined as $\deg(v) := \abs{N(v)}$. 
We include a list of all graphs on three or four vertices with their corresponding names in Figure \ref{fig::graphs}.

\begin{figure}[ht]
   \centering
   \begin{tikzpicture}[scale=1.5, every node/.style={circle, draw, fill=white, inner sep=2pt}]
    % Define the vertices
    \node (v0) at (-0.75,0) {0};
    \node (v1) at (0,0) {1};
    \node (v2) at (0.5,0.5) {2};
    \node (v3) at (1,0) {3};
    \node (v4) at (0.5,-0.5) {4};
    \node (v5) at (1.5,0.5) {5};
    \node (v6) at (1.5,-0.5) {6};
    
    % Draw the edges
    \draw (v0) -- (v1);
    \draw (v1) -- (v2);
    \draw (v1) -- (v3);
    \draw (v2) -- (v3);
    \draw (v2) -- (v4);
    \draw (v3) -- (v5);
    \draw (v4) -- (v6);
    \draw (v5) -- (v6);
    \draw (v1) -- (v4);
\end{tikzpicture}
\caption{An Example Graph}
\label{fig::exampleGraph}
\end{figure}
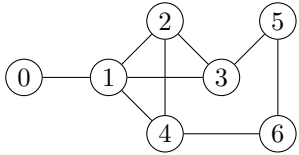

A graph $H$ is an \emph{induced subgraph} of a graph $G$ if there are $\abs{V(H)}$ vertices in $G$ such that, when restricting the edges of $G$ to these vertices, the resulting graph is isomorphic to $H$. 
We refer to $G$ as the \emph{host graph}.
Moving forward, we will simply refer to $H$ as a \emph{subgraph} of $G$, as our results do not concern non-induced subgraphs. 
Define the egocentric-network of a vertex $v$ to be the subgraph induced by $N(v)$. 
The subgraph density of $H$ in $G$ is the number of subgraphs of $G$ isomorphic to $H$, divided by a normalizing factor. 
That is, \[ \frac{\abs{ \{ K \text{ is an induced subgraph of } G : K \cong H \}}} {\dbinom{ \abs{V(G)}}{ \abs{V(H)}}}. \]
For example in figure \ref{fig::exampleGraph} the density of $K_3$ -- the complete graph on three vertices -- is $2 / {6 \choose 3}$.

In this paper, we are primarily interested instead in the density in egocentric networks, that is subgraph densities in $N(v)$. 
For example, the edge density in the egocentric network of vertex $1$ in Figure \ref{fig::exampleGraph} is $2/ {4 \choose 2}$ (edges  $\{2,3\}$ and $\{2,4\}$) whereas for vertex $6$ it is $0/ {2 \choose 2}$. 
If, for example, we were considering the localized point cloud of this graph with respect to $P_3$ (the path of three vertices) vertex $1$ would correspond to the coordinate $\left(2/ {4 \choose 2}, 1/ {4 \choose 3}\right)$. 

\subsection*{An Overview of Flag Algebras}
We provide a high-level overview of the plain flag algebra method \cite{flag}. 
Our discussion focuses on the method in the context of approximations of the feasible region of subgraph densities. 
Refer to any of \cite{thesis, flagIntro, flag} for more thorough background. 

Per convention, we represent subgraph densities pictorially, especially when the host graph's identity is clear or irrelevant.
For instance, $\LittleKtwo$ denotes the probability of selecting two vertices joined by an edge in the host graph.

Razborov's seminal work \cite{flag} defines an algebra over subgraph densities.
This algebra allows us to prove (in)equalities concerning subgraph densities.
For example, one can show that $\LittleKtwo \,+ \LittleKtwoComp \,= 1$ regardless of host graph; or, that \[ \Kthree \,+ \KthreeComp \,\ge 0.25.\]
That is, for any possible graph $G$, the subgraph density of $K_3$ in $G$ plus the subgraph density of $\overline{K_3}$ in $G$ is always at least $0.25$. 
This is asymptotically equivalent to Goodman's theorem \cite{goodman} and can be proved using flag algebras \cite{AxelThesis}.

One notable use of flag algebras is the \emph{plain flag algebra method}.
This method enables one to generate valid algebraic inequalities computationally via semi-definite programming. 
More specifically, it allows one to approximate \emph{graph programs}.
A graph program consists of an objective function of subgraph densities and constraints of subgraph densities. 
For example, one can rephrase Goodman's theorem as: ``the graph program 
\[ \min \Kthree + \KthreeComp \]
 has an objective value of $0.25$''.
Bounds derived via flag algebras are only asymptotically true, i.e., true for host graphs containing arbitrarily many vertices.
This is because the method intrinsically ignores $o(1)$ terms that are negligible in the infinite case.
In finite graphs, this can cause points to lie outside the feasible region (e.g., Figure \ref{algorithmExample}). 
We do not apply existing techniques to account for lower-order terms \cite{C5}, as any points lying outside the feasible region are already classified as outlier in our method. 

In practice, the plain flag algebra method approximations of graph programs are accurate to several decimal places \cite{C5, hyper, crossing}. 
The method is rarely exact, but its accuracy can be improved by increasing a paramter known as \emph{the number of vertices run on} (effectively increasing the size of the semi-definite program), at the expense of an exponential increase in runtime. 
To balance accuracy and reasonable computation time, we run the plain flag algebra method on $6$ vertices.
The accuracy of the plain flag algebra method is discussed in more detail in Section \ref{sec:results}. 

The explicit details of the algorithms used for generating the semi-definite programs can be found in \cite{thesis} which includes algorithms and an implementation in C++.
We do not include any specifics here as that is beyond the scope of this paper.

\section{Algorithm for calculating the subgraph spread ratio.}\label{alg}
The subgraph spread ratio is formally defined by the algorithm detailed below. 
The assumptions made in our algorithm were aimed at creating a ``natural'' and intuitive approach to the problem, but we indicate instances in which other assumptions could be made. 

The central algorithm is divided into four distinct steps: \begin{enumerate}
\item Calculating the feasible region using the plain flag algebra method.
\item Calculating the localized point cloud.
\item Pruning the dataset for outliers.
\item Computing the subgraph spread ratio.
\end{enumerate}

C++ code for computing the subgraph spread ratio can be found at \url{https://github.com/sandialabs/Tight-Practical-Bounds-for-Subgraph-Densities-in-Ego-centric-Networks}.

We refer to Figure \ref{algorithmExample} as a concrete, visual example of the method. 
\subsection*{Calculating the Feasible Region}

We first give some intuition about the feasible region in Figure \ref{algorithmExample} (the shaded region).
For this localized point cloud, the $y$-axis is the density of $\overline{P_3}$, the graph on three vertices with exactly one edge. 
If the edge density is too low or too high, then we must have fewer $\overline{P_3}$ structures, since the graph would contain too many empty or complete subgraphs, respectively. 
Indeed, in this case, it is well-known that the graph containing the most occurrences of $\overline{P_3}$ is two disjoint, balanced, complete graphs \cite{goodman}.
Such a graph asymptotically has edge density $1/2$, which is why we see the peak of our feasible region at $x = 1/2$.  

We next discuss how to approximate the feasible regions using the plain flag algebra method. 
This computation is independent of the dataset and is reused. 

We are interested in solving the graph program 
\begin{equation}\label{graphProgram}
       \max H \text{ subject to } \frac{i}{N} \le \LittleKtwo \le \frac{i+1}{N} \text{ for all } 0 \le i < N,
\end{equation}
where $N$ is a pre-selected level of accuracy and $H$ is an arbitrary subgraph.
In Figure \ref{algorithmExample}, $H = \overline{P_3}$.
For our analysis, we calculate these graph programs for all $H$ on at least $3$ and $4$ vertices, and all $0 \le i < N$ with $N = 100$. 
We choose $N=100$ as it gives a high level of accuracy while still allowing the plain flag algebra method to run in a reasonable amount of time, around $15$ minutes per subgraph on a 2020 MacBook Pro with M1 chip.

Practically, concerning a localized point cloud with respect to $H$, \eqref{graphProgram} calculates the largest our $y$ coordinate ($H$ density) can be when our $x$ coordinate (edge density) is between $i/N$ and $(i+1)/N$.
This creates a step function with $N$ steps. 
Calculating a discrete step function is significantly easier than calculating a continuous feasible region. 
This is analogous to linear programming, which can be solved in polynomial time \cite{LP} and parametric linear programming, which can't \cite{parametric}. 
Although a continuous feasible region has been calculated for complete graphs \cite{K3Dens, K4Dens, KnDens}, these techniques require extremely sophisticated machinery, cannot be automated, and would likely require a significant amount of work to expand to other graphs. 
We don't use these results in this work as the resulting formulas are complex and would be difficult to implement.
Further, as discussed in the Results section, the plain flag algebra method produces results which are sufficiently accurate. 

In $\eqref{graphProgram}$ we only calculate the upper bound of our feasible region because, when $H$ is not a complete or empty graph, the lower bound of the feasible region is $0$ \cite{main} (see for instance Figure \ref{allFigs}). 
For $H = K_n$ or $\overline{K_n}$ we do solve  \eqref{graphProgram} as a minimization problem as well. 

We make one exception to using the plain flag algebra method for calculating the feasible region. 
When upper bounding the feasible region for complete and empty graph, we instead use the Kruskal-Katona theorem \cite{kruskal, katona}. 
In particular, this theorem tells us that
\[ K_n \le \Ktwo^{\,n/2}, \text{ and} \]
\[ \overline{K_n} \le \left(1-\Ktwo\right)^{\,n/2}. \]
We use these bounds as opposed to the bounds from the plain flag algebra method as they are tighter than the plain flag algebra method and are computationally easily expressible. 

\subsection*{Calculating the Localized Point Cloud}
In this step, we determine the subgraph densities in the neighborhood of each vertex. 
This yields the scatter plot of points (see Figure \ref{algorithmExample}) referred to as the \emph{localized point cloud} with respect to $H$, where $H$ is the subgraph in \eqref{graphProgram}. 
For this computation, we iterate through each vertex in the graph and then calculate the density of each subgraph in its neighborhood using ESCAPE \cite{escape}. 
To reduce the impact of $o(1)$-order errors generated by the plain flag algebra method, we restrict our analysis to vertices with degree of at least $10$. 
We see in Figure \ref{algorithmExample}, that while there are still some vertices that fall outside the asymptotic feasible region, none fall far outside it. 

Subgraph densities are used due to convention, simplicity, and compatibility with flag algebras. 
Other meaningful measures like homomorphism density could be considered in future work.

\subsection*{Pruning the Data Set}
Our next step is to determine the outliers in localized point cloud, which are shown in red in Figure \ref{algorithmExample}. 
To do this, we identify an area of the localized point cloud containing the majority of points and exclude any points outside this area. 
Specifically, we employ a method inspired by topological data analysis and, particularly, persistent homology. 
Persistent homology investigates the $k$-dimensional holes (connected components, loops, voids, etc.) generated by connecting points in a point cloud according to some distance metric \cite{TDA}. 
While we do not employ any notion of \textit{persistence} herein, we use the technique of identifying connected components within a distance threshold for our analysis. 

More specifically, We first create balls of equal radius around each data point. 
Then the radius is iteratively adjusted until one connected area contains a fixed percentage of points.
We select $95\%$ of the points as non-outliers and have our radius accurate to six decimal points.
An exact description of the algorithm as psuedocode is included below.

Let $D$ be the $n \times n$ matrix such that $D[i,j]$ is the distance between vertices $i,j$ in the localized point cloud. 
In this algorithm, we compute the largest connected component of a graph. 
This can be done quickly using a depth-first search algorithm. We do not include the specifics of that step.
For a $0-1$ symmetric matrix, $A$, with $0$'s on the diagonal, define $G[A]$ to be the graph with adjacency matrix $A$. 

\begin{algorithm}[H]
   \caption{Pruning Algorithm}
   \begin{algorithmic}
          \State $\alpha = 0.95$ \Comment Percentage of points we wish to keep.
          \State $\epsilon = 10^{-6}$ \Comment Tolerance on diameter of balls.
          \State $lb = 0$ \Comment Lower bound on diameter of balls.
          \State $ub = 1$ \Comment Upper bound on diameter of balls.

          \While{$ub - lb > \epsilon$}
                 \For{$i,j \in [n]$} \Comment Create a graph where vertices are connected if they are `close'.
                        \If{$D[i,j] < (ub+lb)/2$} 
                               \State $A[i,j] = 1$.
                        \Else
                               \State $A[i,j] = 0$.
                        \EndIf
                 \EndFor

                 \State Set $X$ to be the largest connected component in $G[A]$.

                 \If{ $\frac{\abs{X}}{n} < \alpha$ }
                       \State $lb = (ub+lb)/2$.
                 \Else
                        \State $ub = (ub+lb)/2$.
                 \EndIf
          \EndWhile

          \State Return $X$.
   \end{algorithmic}
\end{algorithm}

In order for this area to be meaningful, we require that the points approximate a single connected region. 
Although there is no reason apriori for this to be true, it does appear to be the case for the data we used. 
This algorithm could be adjusted to account for multiple regions if necessary.

\subsection*{Calculating the subgraph spread ratio}

We compute the approximate area occupied by the localized point cloud and compare it to the area of the feasible region. 
Calculating the latter is a simple summation over the upper and lower bounds provided by the plain flag algebra method. 
As for calculating the ``area'' of the localized point cloud, the difficult work is done when pruning the data set, as the union of the balls provides an estimate of the area covered by the points.
We approximate this area with a Monte Carlo sampling of one million points. 
By the Chernoff bound, this gives three decimal places of accuracy for our calculations of subgraph spread ratios.
We use a Monte Carlo simulation for its ease of implementation and ability to be tuned for accuracy, but exact area computations are certainly possible, by using, for example, the principle of inclusion, exclusion. 
Finally, the ratio of this area to the area of the feasible region is the subgraph spread ratio.
In Figure \ref{algorithmExample}, the subgraph spread ratio is around $0.4$. 

In cases where the pruned localized point cloud does not occupy the entire domain, we restrict the area of the feasible region to the corresponding subdomain, as in Figure \ref{allFigs2} for the point cloud with respect to $K_4 - e$. 
This ensures that we do not underestimate the subgraph spread ratio in such cases. 

\section{Experimental Results \label{sec:results}}

%\subsection*{Data Selection, Assumptions, and Limitations}

The datasets referenced in this study are sourced from the Stanford Large Network Dataset Collection (SNAP) \cite{snap}. 
We did not analyze any directed graphs, as such analysis is better performed with methods specific to directedness and could be pursued in future work.
Further, we restricted ourselves to graphs with at most five million edges for computational feasibility.

Our analysis assumes a full range of edge densities in each vertex's neighborhood. 
We enforce that the unpruned localized point clouds cover the x-axis. 
This restriction prevents skewing our subgraph spread ratios by systematically looking at only a portion of the domain. 
Subsets of the domain may not accurately approximate the subgraph spread ratio across the entire domain.

Our empirical study included eight Facebook graphs from GEMSEC \cite{GEMSEC} and MUSAE \cite{MUSAE} and three from Wikipedia links from MUSAE \cite{MUSAE}. 
The nodes in the Facebook data are pages, with edges between pages if at least one person likes both pages. 
The MUSAE Facebook data is from various page types, while each GEMSEC dataset is topic-specific, such as TV shows or politicians. 
There may be overlap between the Facebook data from the two sources, as the lead authors in \cite{GEMSEC, MUSAE} are the same. 
The nodes in the Wikipedia data are pages linked to a specific page (Crocodile, Squirrel, and Chameleon), with edges between pages sharing a common link. 
These graphs range from approximately $\num{4000}$ vertices and $\num{17000}$ edges to $\num{50000}$ vertices and $\num{800000}$ edges. 
Our algorithm ran in around $30$ seconds for the smallest graph and just under an hour for the largest, on a 2020 MacBook Pro with M1 chip. 
These algorithms are not optimal and run-times could be decreased.

We begin with empirical results from two specific datasets: a subset of Facebook pages \cite{MUSAE} and links in the "Crocodiles" Wikipedia article \cite{MUSAE}. 
We then extend our analysis to the entire dataset. 
Finally, we address the theoretic results of our analysis, specifically the improved bounds on the feasible region estimate, using evidence from extremal graph theory.

In this discussion, we often conjecture why we observe certain subgraph spread ratios and localized point clouds. 
Our goal is not to provide definitive explanations but to convey the meaning that the subgraph spread ratio captures about structures in real-world networks. 
While no singular graph-theoretic measure can fully explain a specific network structure, the subgraph spread ratio can inform mathematically meaningful conjectures and enable relevant comparisons of these complex objects. 
Ultimately, we wish to demonstrate the practical utility of the subgraph spread ratio for both mathematicians and domain experts in applied fields.

\subsection*{Comparisons of Subgraph Spread Ratios Across Two Data Sets}

\begin{figure*}[h]
      \centering
       \includegraphics[scale=0.09]{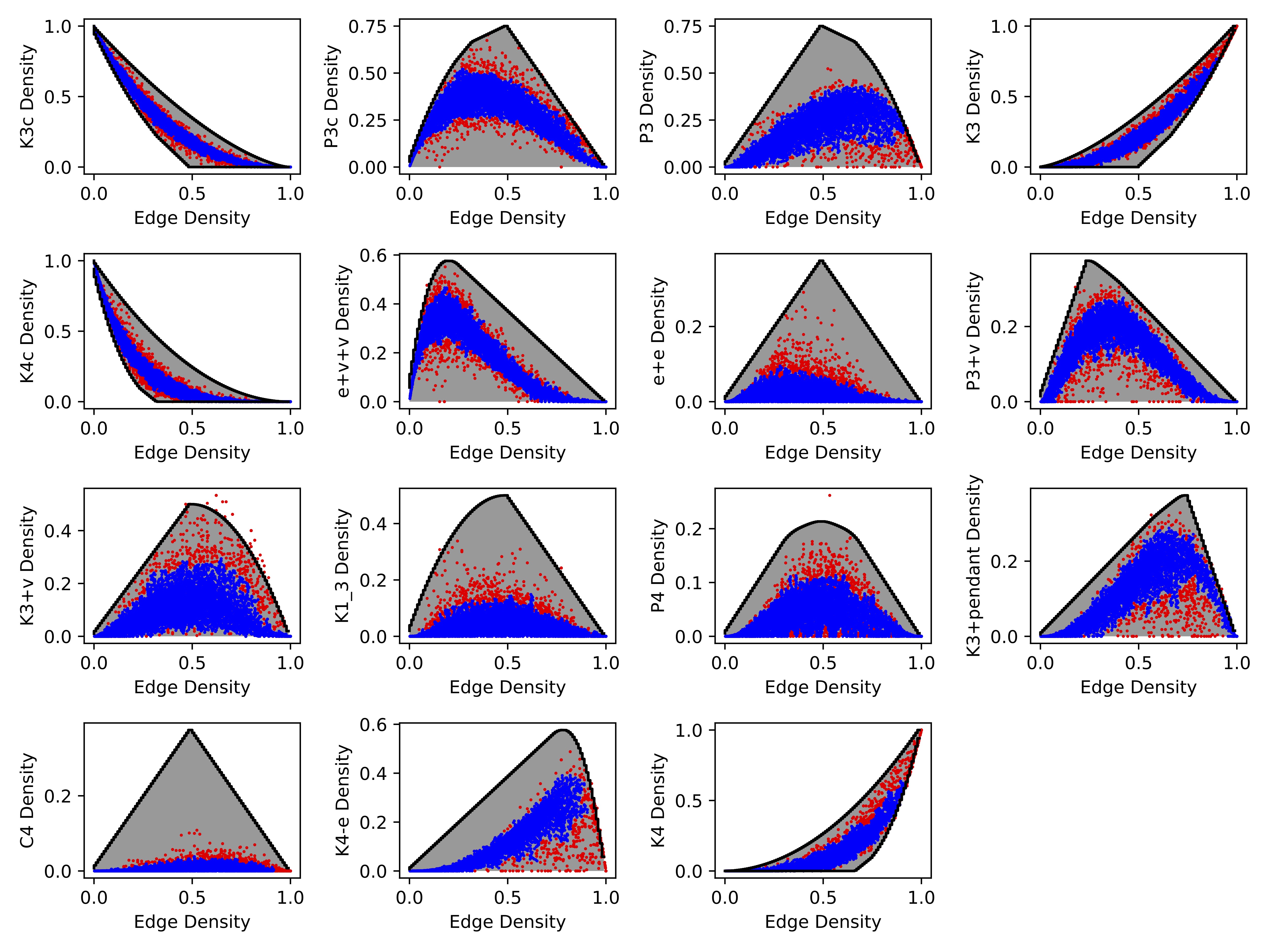}
       \caption{Plot of all localized point clouds of $H$ for $\abs{V(H)} \le 4$ of Facebook data \cite{MUSAE}.
       The gray region is the feasible region and red data points are outliers. 
       A list of subgraph spread ratios can be found in Tables \ref{myTable1} and \ref{myTable2} under FB, the row is highlighted. 
       The y axes follow the same layout as in Figure \ref{fig::graphs}. 
       }
       \label{allFigs}
\end{figure*}

\begin{figure*}[h]
   \centering
       \includegraphics[scale=0.09]{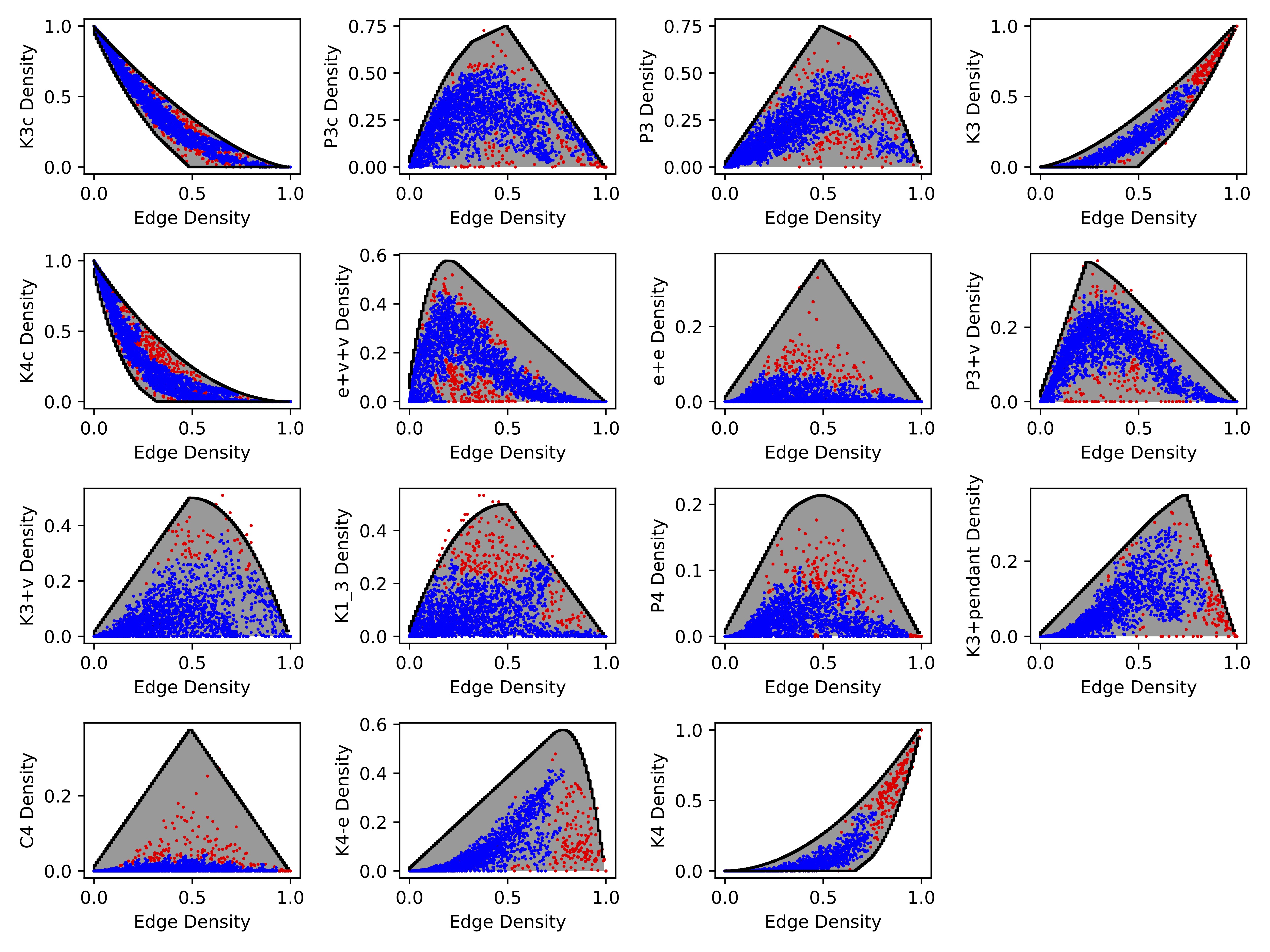}
       \caption{Plot of all localized point clouds of $H$ for $\abs{V(H)} \le 4$ Wikipedia links \cite{MUSAE} on the Crocodile page.
       The gray region is the feasible region and red data points are outliers. 
       A list of subgraph spread ratios can be found in Tables \ref{myTable1} and \ref{myTable2} under Wiki Crocodile, the row is highlighted. 
       The y axes follow the same layout as in Figure \ref{fig::graphs}. 
       }
       \label{allFigs2}
\end{figure*}

We begin with a discussion of two specific networks: the MUSAE Facebook page data and the Wikipedia link data for the Crocodile page \cite{MUSAE}.
These networks have approximately $\num{22000}$ and $\num{11000}$ nodes respectively and both have approximately $\num{170000}$ edges. 
We choose these two networks as the come from the same source and have relatively large number of nodes and edges. 
We focus on two specific graphs to begin with as this allows us to make concrete comparisons and conjectures based on specific localized point clouds and subgraph spread ratios. 
Plots of the localized point clouds can be found in Figures \ref{allFigs} and \ref{allFigs2} respectively and a list of their subgraph spread ratios can be found in Tables \ref{myTable1} and \ref{myTable2} in the highlighted rows. 

The subgraph spread ratio allows us to classify and compare networks.
The feasible region quantifies the range of features that are possible for an arbitrary graph and the subgraph spread ratio measures how much of that region is realized by a given graph. 
The Facebook page dataset has, on average, an subgraph spread ratio of $0.38$ while the Wikipedia dataset has that of $0.52$. 
This indicates that the Wikipedia dataset contains more of the potential subgraph structures.
By contrast, the Facebook dataset contains a smaller proportion of all possible subgraph densities, implying that there are more extrinsic factors driving its structure.
Thus the average subgraph spread ratio quantifies our intuitive understanding that social networks, such as Facebook pages, are structurally distinct from link graphs. 

Examining the subgraph spread ratio in relation to specific subgraph structures can provide additional insight to community structure networks. 
Figures \ref{allFigs} and \ref{allFigs2} contain the $15$ localized point clouds of subgraphs of size up to $4$ for the Facebook and Wikipedia networks.
We highlight the striking difference in both networks through the localized point clouds with respect to $\overline{P_3}$, the graph on three vertices with a single edge. 
The subgraph spread ratio of $\overline{P_3}$ for the Facebook data is $0.39$, whereas the corresponding subgraph spread ratio for the Wikipedia data is $0.69$. 
Furthermore, we note that the localized point cloud for the Facebook data also looks notably different than for Wikipedia data. 
For Facebook data, there is a large region of low density of $\overline{P_3}$ which does not occur, but which does occur however in the Wikipedia data.
This is despite the fact that the Facebook data has twice as many nodes as the Wikipedia data.
One possible explanation of this is because of the community structures of the respective graphs. 
We know that $\overline{P_3}$ occurs in the Facebook network when pages for two topics, $t_1$ and $t_2$, share at least one member with each other and none with a third topic $t_3$. 
So, if for example, $t_1$ and $t_2$ were part of one community while $t_3$ was part of a separate community, this would create an $\overline{P_3}$, leading to the higher observed density.
In contrast, the Wikipedia pages may not have as high of occurrences of $\overline{P_3}$ since two Wikipedia pages $w_1, w_2$ that are linked by a topic will likely both link to a third topic $w_3$, since all three pages are within a community of similar concepts. 
Thus, the localized point clouds with respect to $\overline{P_3}$ highlight a difference in the community structure of these two datasets. 

We compute the subgraph spread ratios for all additional subgraphs on four or fewer vertices and include the results in Tables \ref{myTable1} and \ref{myTable2} for all of our data. 
While there are notable differences between the Wikipedia graph and Facebook page graph, particularly $\overline{K_3}, \overline{P_3}, \overline{K_4}, e+v+v$ and $K_{1,3}$, there are other subgraphs for which the differences between subgraph spread ratios are relatively similar. 
This negative result demonstrates that not all subgraph structure is meaningful for all graphs. 
Indeed, the question of identifying which subgraph structures are important for a specific network would be of great interest to domain experts, since it may indicate unique phenomena inherent to the dataset.
Such results may also inform the study of graph generation, particularly in comparing the community structure of artificial and real graphs, for model validation.

% First part of the table
\begin{table*}[h]
   \centering
   \caption{Subgraph spread ratios for some networks in SNAP \cite{snap}. 
   Rows are sorted by average subgraph spread ratio, similar to the bar chart. 
   The rows of data from Figures \ref{allFigs} and \ref{allFigs2} are highlighted.
   This table only contains three vertex subgraphs.}
   \label{myTable1}
   \scalebox{0.75}{
   \begin{tabular}{|c||c|c|c|c|}
   \hline
   & $\overline{K_3}$ & $\overline{P_3}$ & $P_3$ & $K_3$ \\
   \hline
   \hline 
   FB Politician \cite{GEMSEC} & 0.425 & 0.392 & 0.379 & 0.375 \\
   \hline
   FB Public Figure \cite{GEMSEC} & 0.355 & 0.339 & 0.400 & 0.606 \\
   \hline
   FB Government \cite{GEMSEC} & 0.410 & 0.387 & 0.408 & 0.442 \\
   \hline  
   \rowcolor{yellow} FB \cite{MUSAE} & 0.355 & 0.393 & 0.483 & 0.498 \\
   \hline 
   FB New Sites \cite{GEMSEC} & 0.368 & 0.432 & 0.463 & 0.477 \\
   \hline
   FB Athletes \cite{GEMSEC} & 0.403 & 0.479 & 0.456 & 0.531 \\
   \hline 
   FB Company \cite{GEMSEC} & 0.395 & 0.576 & 0.629 & 0.700 \\
   \hline 
   FB TV Show \cite{GEMSEC} & 0.495 & 0.459 & 0.560 & 0.695 \\
   \hline
   \rowcolor{yellow} Wiki Crocodile \cite{MUSAE} & 0.636 & 0.686 & 0.556 & 0.621 \\
   \hline
   Wiki Squirrel \cite{MUSAE} & 0.644 & 0.660 & 0.571 & 0.679 \\
   \hline 
   Wiki Chameleon \cite{MUSAE} & 0.811 & 0.751 & 0.676 & 0.833 \\
   \hline
   \end{tabular}
   }
\end{table*}

% Second part of the table
\begin{table*}[h]
   \centering
   \caption{Table \ref{myTable1} continued. Subgraph spread ratios for four vertex subgraphs. 
   Average is across both three and four vertex subgraphs.}
   \label{myTable2}
   \scalebox{0.75}{
   \begin{tabular}{|c||c|c|c|c|c|c|c|c|c|c|c|c|}
   \hline
   & $\overline{K_4}$ & $e+v+v$ & $e+e$ & $P_3 + v$ & $K_3 + v$ & $K_{1,3}$ & $P_4$ & $K_3+$pendant & $C_4$ & $K_4-e$ & $K_4$ & Average \\
   \hline
   \hline 
   FB Politician \cite{GEMSEC} & 0.392 & 0.320 & 0.197 & 0.447 & 0.402 & 0.289 & 0.484 & 0.423 & 0.111 & 0.235 & 0.310 & 0.345 \\
   \hline
   FB Public Figure \cite{GEMSEC} & 0.391 & 0.367 & 0.172 & 0.474 & 0.430 & 0.247 & 0.460 & 0.405 & 0.093 & 0.173 & 0.412 & 0.355 \\
   \hline
   FB Government \cite{GEMSEC} & 0.336 & 0.277 & 0.220 & 0.446 & 0.450 & 0.258 & 0.454 & 0.489 & 0.090 & 0.260 & 0.422 & 0.357 \\
   \hline  
   \rowcolor{yellow} FB \cite{MUSAE} & 0.328 & 0.321 & 0.198 & 0.443 & 0.544 & 0.242 & 0.429 & 0.525 & 0.077 & 0.359 & 0.491 & 0.379 \\
   \hline 
   FB New Sites \cite{GEMSEC} & 0.357 & 0.370 & 0.195 & 0.514 & 0.637 & 0.271 & 0.394 & 0.527 & 0.088 & 0.276 & 0.447 & 0.388 \\
   \hline
   FB Athletes \cite{GEMSEC} & 0.389 & 0.403 & 0.221 & 0.521 & 0.658 & 0.276 & 0.391 & 0.636 & 0.104 & 0.319 & 0.418 & 0.414 \\
   \hline 
   FB Company \cite{GEMSEC} & 0.380 & 0.413 & 0.137 & 0.587 & 0.528 & 0.321 & 0.381 & 0.665 & 0.137 & 0.432 & 0.604 & 0.459 \\
   \hline 
   FB TV Show \cite{GEMSEC} & 0.411 & 0.360 & 0.202 & 0.555 & 0.613 & 0.276 & 0.409 & 0.662 & 0.089 & 0.491 & 0.705 & 0.465 \\
   \hline
   \rowcolor{yellow} Wiki Crocodile \cite{MUSAE} & 0.619 & 0.515 & 0.200 & 0.617 & 0.654 & 0.511 & 0.315 & 0.649 & 0.082 & 0.512 & 0.585 & 0.517 \\
   \hline
   Wiki Squirrel \cite{MUSAE} & 0.673 & 0.548 & 0.229 & 0.650 & 0.629 & 0.456 & 0.344 & 0.622 & 0.080 & 0.474 & 0.549 & 0.521 \\
   \hline 
   Wiki Chameleon \cite{MUSAE} & 0.797 & 0.485 & 0.233 & 0.685 & 0.560 & 0.614 & 0.447 & 0.695 & 0.160 & 0.525 & 0.825 & 0.607 \\
   \hline
   \end{tabular}
   }
\end{table*}

\subsection*{Comparisons of the Subgraph Spread Ratios Across All Data Sets}

\begin{figure}
   \centering
       \includegraphics[scale=0.2]{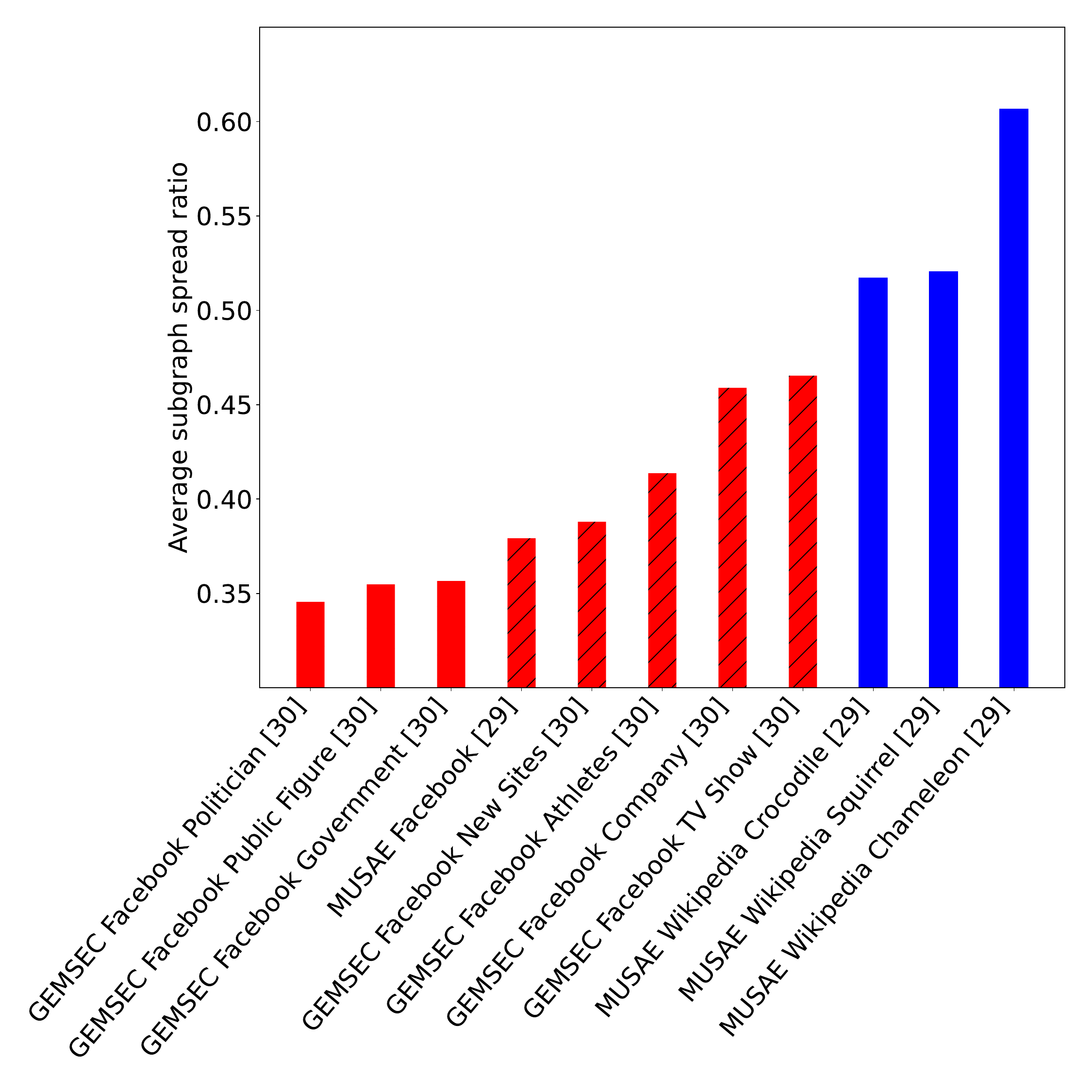}
       \caption{A bar chart comparing the averaged subgraph spread ratios of different networks. 
       Red, striped bars represent social networks, blue un-striped represent Wikipedia links. 
       Entries are sorted by average subgraph spread ratio.}
       \label{barChart}
\end{figure}

Similar results also hold across our entire set of networks.
Figure \ref{barChart} shows the averaged subgraph spread ratio across our 11 distinct networks. 
Note that all the social networks have a lower average subgraph spread ratio than the link graphs, further substantiating that the subgraph spread ratio quantifies informative structural differences between such graphs.
%This figure also demonstrates a key property of the subgraph spread ratio: its robustness to changes in scale. 
%As described previously, these social networks (drawn from the GEMSEC dataset) are possibly subsets of the larger MUSAE Facebook dataset and we would expect their subgraph spread ratio to be reflective of that. 
%The average subgraph spread ratio for the MUSAE Facebook data fall within the result of the average values from the GEMSEC networks, suggesting that the subgraph spread ratio is not sensitive to subcategories within the same network. 
%The ability to study smaller sub-networks enables us to examine networks of immense scale, since we can examine subsets of it without significantly affecting the subgraph spread ratio.  

The spread of subgraph spread ratios for the GEMSEC Facebook pages is also worth examining.
Typical variations in datasets contribute to this spread, but they are unlikely to be the only causes.
We expect Facebook pages of different topics to experience different extrinsic pressures.
For instance, pages for local politicians may have membership siloed by personal politics or geography, unlike pages for TV shows.
This is reflected in the data: Facebook politician pages have the lowest average subgraph spread ratio, while TV show pages have the highest.
The subgraph spread ratio quantifies the extent to which subgraph structure in a network is extrinsically defined and allows us to differentiate the extent to which different networks may be affected by known extrinsic forces.

The range of subgraph spread ratios across different subgraphs for a single network can also be informative.
Consider any of the social networks in this study.
The $C_4$ subgraph spread ratio is consistently small for such networks.
This may be because the presence of $C_4$ subgraphs necessitates the presence of complete bipartite graphs \cite{fourInd} which in turn necessitate induced $P_3$'s.
However, it is well known \cite{social} that $P_3$'s do not frequently appear in social networks.
On the other hand, the subgraph spread ratios for complete graphs are comparatively large, indicating fewer social pressures restricting clique densities.
This example highlights how the subgraph spread ratio reproduces known facts about social networks.

As a final note, we remark on the symmetry between the subgraph spread ratio for $H$ and the subgraph spread ratio for $\overline{H}$.
Such symmetry does not hold for networks observed globally but does hold for egocentric networks.
For example, the subgraph spread ratio for $C_4$ and the subgraph spread ratio for $e+e$ are consistently the smallest across all our data.
This symmetry, though less striking, holds across other subgraphs and their complements and can be observed in the localized point clouds.
If edges and non-edges (corresponding to $H$ and $\overline{H}$) are the "same," the localized point clouds roughly mirror each other.
Within an egocentric network, edges and non-edges behave more similarly, even if this is not the case in the entire graph.

\subsection*{Accuracy of the Plain Flag Algebra Method}

Our novel application of the plain flag algebra method allows us to more accurately characterize the feasible region. 
The theory of inducibility from extremal graph theory indicates that the bounds we achieve using this method are indeed nearly tight. 
Inducibility is defined as the objective value of the graph program $\max H$ for some graph $H$.
Equivalently, it is the maximum induced subgraph density possible for $H$ \cite{thesis}.
The inducibility of $H$ is then a lower bound for the maximum of the feasible region of $H$. 
Therefore, the inducibility, which has been derived theoretically for any graph on at most $4$ vertices, except for the path on $4$ vertices \cite{fourInd}, provides a statistic against which to compare our feasible region estimate.
We find that for any graph, excepting $P_4$, the maximal value of the feasible region is exactly the known inducibility. 
That is, for these graphs, the plain flag algebra method produces a tight upper bound on the feasible region. 
In the case of $P_4$, our computed maximum is $\approx 0.214$, whereas the inducibility is known to be between $0.2014$ \cite{fourInd} and $0.204513$ \cite{flagmatic}.  
While these results do not guarantee the tightness of the plain flag algebra bound not at the maximum, they are a strong indicator of the performance of this method. 

Further, we see in Figures \ref{allFigs} and \ref{allFigs2} that, excluding a few exceptions, there are data points that occur close to the computed boundary. 
This would not occur if the feasible region computations were loose, in which case, none of the localized point cloud would fall near the boundary.
A notable exception is the $C_4$ density on social network data (Figure \ref{allFigs}). 
However, as previously noted, there are legitimate social reasons why we would not expect to see many $C_4$s.

\section{Future Work}\label{sec:future}

This work introduces the subgraph spread ratio and explores its preliminary applications to real-world networks. 
Future research is needed to examine the subgraph spread ratio across varied networks to understand its significance. 
The subgraph spread ratio could become a crucial tool for quantifying structural characteristics in large networks, facilitating comparisons across multiple scales. 
It can also enhance our understanding of real-world networks and improve graph generation for modeling purposes. 
We hope these results will encourage domain experts to apply the subgraph spread ratio to their network data.

Extending our analysis to other network types, such as directed graphs, hypergraphs, or graphs with labeled edges, is an important next step. 
Directed graphs, for instance, capture additional information in social networks or email communications \cite{snap}. 
While our current analysis could be naively applied by ignoring edge direction, this approach sacrifices valuable information. 
The plain flag algebra method has been shown to work on these structures and any general combinatorial structure \cite{directed, labeledEdge, hyper, flag}. 
Therfore, an extension of this algorithm could enable subgraph spread ratio calculations for such networks.

We make several modeling assumptions throughout this work. 
Exploring alternative assumptions could yield valuable insights and open new research avenues. 
Notably, we assume the localized point cloud exists only in $\mathbb{R}^2$, with edge density as the independent variable. 
Other statistics, such as $P_3$ density, could also be meaningful for comparison, particularly in social networks \cite{social}. 
Comparing multiple statistics could create a higher-dimensional point cloud, as visualized in \cite{main}, but many questions remain about analyzing these higher-dimensional clouds.

Our focus has been on holistic graph analysis, but the localized point cloud also provides information about individual vertices. 
For example, Figure \ref{allFigs} shows a page with significantly higher P4 density and a cluster of points with unexpectedly high $C_4$ density. 
Identifying abnormal Facebook pages, such as those spreading fake news, is a pertinent issue that these techniques could help address.

\section{Conclusion}\label{sec:conclusion}

In the study of social networks, a fundamental question arises: are the observed structures driven by social factors or do they arise because of mathematical properties? 
This question is particularly pertinent when examining ego-centric networks and comparing subgraph densities. As shown, for example, in Figure \ref{allFigs}, there are often striking structures in the localized point cloud.

Our contributions in this paper are three-fold. 
First, although there have been attempts to classify the feasible region of the localized point cloud, they have been inadequate when compared to known results from extremal graph theory. 
We use the plain flag algebra method to create tighter feasible regions. 
Using both evidence from extremal graph theory and the data itself, we conclude that these feasible regions are quite accurate.

Secondly, leveraging these improved bounds, we introduce the subgraph spread ratio, as a novel metric. 
This subgraph spread ratio provides a quantitative measure of how much of the structure we see in networks is from mathematically-determined reasons and how much is from domain-driven reasons. 
Mathematically, the subgraph spread ratio measures the percentage of the feasible region that our localized point cloud fills, using techniques inspired by topological data analysis.

Finally, we compute the subgraph spread ratio for various real-world graphs. 
Our data shows that social networks consistently have smaller subgraph spread ratios than networks of links, such as those found in Wikipedia page linkage networks. 
These findings demonstrate the subgraph spread ratio's utility as a robust statistic for network analysis, encapsulating intuitive aspects of social network structures. 
Additionally, our analysis reveals that examining the subgraph spread ratios across various subgraphs can yield insights into the underlying community structures within networks.

\clearpage
\bibliographystyle{plain}
\bibliography{bibliography}

\end{document}